\documentclass[11pt,dvips]{article}

\usepackage{epsfig,times} 
\usepackage{picinpar}
\usepackage{wrapfig}
\usepackage{floatflt}
\makeatletter
\renewcommand{\section}{\@startsection{section}{1}{0in}
        {0.4\baselineskip}{0.1\baselineskip}{\Large\bf}}
\renewcommand{\subsection}{\@startsection{subsection}{2}{0in}
        {0.25\baselineskip}{-\baselineskip}{\large\bf}}
\renewcommand{\subsubsection}{\@startsection{subsubsection}{3}{0in}
        {0.1\baselineskip}{-\baselineskip}{\normalsize\bf}}
\makeatother
\begin{document}

%
%  Session and Paper Code:
%\thispagestyle{myheadings}
%
%  ***INSTRUCTIONS:***  Replace `OG 9.9.9' in the command argument below
%                       with your assigned session and paper code:
%\markright{HE 4.2.03}
%
\makeatletter\newcommand{\ps@icrc}{
\renewcommand{\@oddhead}{\slshape{HE.4.2.03}\hfil}}
\makeatother\thispagestyle{icrc}

\begin{center}
%
%  ***INSTRUCTIONS:***  Replace `Instructions for Preparation of Manuscript'
%                       with your paper's title:
{\LARGE \bf MACRO as a Telescope for Neutrino Astronomy}
\end{center}

%  Author List:
\begin{center}
%
%  ***INSTRUCTIONS:***  Replace authors and addresses below with your own:
%
{\bf T. Montaruli$^{1}$ for the MACRO Collaboration}\\
{\it $^{1}$Istituto Nazionale di Fisica Nucleare, Bari, I-70126, Italy\\}
\end{center}

%  Abstract:
\begin{center}
{\large \bf Abstract\\}
\end{center}
\vspace{-0.5ex}
%
%  ***INSTRUCTIONS:***  Replace text below with your own abstract:
%
We use a sample of 990 upward-going muons,
induced primarily by atmospheric neutrinos, to search for neutrinos of
astrophysical origin.
No evidence has been found using the event direction information. 
Flux limits of the order of $10^{-15}$ 
cm$^{-2}$ s$^{-1}$ are imposed on current models for candidate point-sources.
A space-time correlation search has been undertaken between 2328 BATSE
gamma ray bursts (GRBs) and MACRO upward-going muons.
%
%  Leave this line skip in place:
\vspace{1ex}

%
%  Manuscript text:
%
\section{Introduction: Neutrino Astronomy:}
\label{sec:intro}
\begin{tabwindow}[1,r,%
{\mbox{
\begin{tabular}{|ccc|}\hline
$\cos\theta$ & $\gamma = 2$ & $\gamma = 2.2$ \\ \hline
0.15 & 0.77 & 0.72 \\
0.35 & 0.90 & 0.85 \\
0.55 & 0.91 & 0.87 \\
0.75 & 0.91 & 0.87 \\
0.95 & 0.91 & 0.87 \\ \hline 
\end{tabular}
%\label{tab1} 
}},%
{Fraction of events accepted in a $3^{\circ}$ cone for various  
zenith angles and 2 spectral indices.}]
The detection of high energy ($E_{\nu} > 100$ MeV)  
neutrinos of astrophysical origin would open an exciting field
complementary to $\gamma$-ray astronomy due to their penetrating power. 
TeV $\gamma$-rays suffer absorption in intergalactic 
space on infrared light, PeV $\gamma$s on the microwave background and EeV 
$\gamma$s on radio-waves. 
Astrophysical source models in which radiation is due 
to electromagnetic processes do not produce significant $\nu$ fluxes.
Sources involving ``astrophysical beam dumps'' (accelerated protons
on a gas of matter or radiation with consequent production of
neutrinos from $\pi^{\pm}$ decay) (Gaisser, Halzen \& Stanev, 1995) are 
candidate $\nu$ sources. Examples are supernova remnants (SNRs) in which
particles can interact with the gas in the acceleration region, and X-ray
binaries made of a compact object (neutron star or black hole) and a
non compact companion transferring mass to the other with 
consequent development of an accretion disk (Gaisser, 1996). 
The $\nu$ spectrum expected from ``beam dump'' sources is a power law
with spectral differential index $\gamma \sim 2 - 2.5$ as  
expected from Fermi acceleration mechanisms. 
Neglecting $\gamma$ absorption, $\nu$ fluxes are expected 
almost equal to $\gamma$-ray ones. Hence spectra measured by $\gamma$-ray 
experiments can be used to calculate expected rates of $\nu$ events.
In MACRO they are $\sim 10^{-3}-10^{-2}$ events/yr for a typical point-like 
source.   
The scenario in Protheroe, Bednarek \& Luo, 1998 
for young SNRs, in which particle acceleration takes place 
along the magnetic field ($10^{12}$ Gauss) lines of a 
pulsar of 5 msec period formed during the supernova explosion, would produce 
$\sim 5$ events/yr in $10^3$ m$^2$ 
for $E_{\nu} \ge 100$ GeV during the first 0.1 year after the 
explosion for a beaming solid angle of 1 sr. However the event rate
drops rapidly after 0.1 years. 
\end{tabwindow}
We also explore the possibility suggested by fireball models that GRBs
are $\nu$ sources.
GRBs are interesting as they are the  
most powerful objects ever observed in our Universe. For example,
the total energy of GRB 980329 for isotropic emission is
$> 10^{54}$ erg (Galama et al., 1999).
Predicted rates are 
extremely low for detectors of the size of MACRO and the sensitivity 
needed could probably be reached 
only by a new generation of experiments of huge areas employing
an active medium of sea water or ice.
Nevertheless, MACRO is monitoring almost online the 
available region of sky using the detected $\nu$ events. The absence
of an excess of these events with respect to the fluctuations of the
background of atmospheric $\nu$ events allows us to derive 
flux limits that constrain some theoretical models.
\section{Search for Point-like Sources:}
\label{sec:point}
The MACRO detector is located in the underground Gran Sasso Laboratories
and has dimensions of $12 \times 76.6$ m$^2$ and a height of 9 m. MACRO 
measures $\nu$-induced upward-going muons using a system of 12 m long 
boxes containing $\sim 600$ tons of 
liquid scintillator to  measure the time-of-flight (time resolution
$\sim 500$ psec) and $\sim 20,000$ m$^2$ of streamer
tubes for tracking 
(angular resolution better than 1$^{\circ}$ and pointing accuracy
checked with the moon shadow measurement (Ambrosio et al., 1999)).
The rock absorber inside the lower half 
of MACRO imposes an energy threshold to 
vertical muons of $\sim 1$ GeV.
The data used for the upward-going muon analysis has been collected 
since Mar. 89 with the incomplete detector
(Ahlen et al., 1995); since Apr. 94 the full detector has been taking 
data (Ambrosio et al., 1998).     
In addition to $\sim 33 \cdot 10^{6}$ 
atmospheric 
$\mu$s,
990 upwardgoing $\mu$s with $-1.25 < 1/\beta < -0.75$
are selected with an automated analysis. $1/\beta =  
\Delta T c/L$, $\Delta T$ being the measured T.o.F. and $L$ the track 
length, is $\sim 1$ for downward-going muons and $\sim -1$ for upward-going
muons. 
Among these 990 events, 890 are measured with the full detector. 
The T.o.F. measurement is used to select upward-going
$\mu$s produced in the rock below and inside the apparatus 
by atmospheric neutrinos of average energy $\langle E_{\nu} \rangle \sim 100$ 
GeV and $\langle E_{\nu} \rangle \sim 4$ GeV, respectively, from 
atmospheric downward-going muons. 
The main requirement to reject
events with incorrect $\beta$ measurement is that the position along 
the scintillation boxes measured using the times at the 2 ends
(spatial resolution $\sim 11$ cm) and the position obtained using 
the streamer track (spatial resolution of $\sim 1$ cm) are in agreement 
within 70 cm. 
\begin{table}
\begin{center}
\begin{tabular}{|cccccccc|}\hline
Source& $\delta$& Data ($3^{o}$)&  Backg.($3^{o}$) &  $\mu$-Flux
& $\mu$-Flux & Prev. best& $\nu$-Flux\\ 
  & & & & 
 limit 1 & limit 2 & $\mu$ limit & limit\\ 
      &   & & &cm$^{-2}$ s$^{-1}$& cm$^{-2}$ s$^{-1}$ 
& cm$^{-2}$ s$^{-1}$ & cm$^{-2}$ s$^{-1}$ \\ \hline
 SMCX-1    & -73.5$^{\circ}$ & 3 & 1.87 & $0.60 \cdot 10^{-14}$ 
&$0.67 \cdot 10^{-14}$ & - & $0.19 \cdot 10^{-5}$\\    
 SN1987A   & -69.3$^{\circ}$ & 0 & 1.79 & $0.29 \cdot 10^{-14}$ 
&$0.16 \cdot 10^{-14}$ & $1.15 \cdot 10^{-14}$ B& $0.09 \cdot 10^{-5}$\\
 Vela P& -45.2$^{\circ}$ & 1 & 1.40 & $0.56 \cdot 10^{-14}$ 
&$0.53 \cdot 10^{-14}$  & $0.78 \cdot 10^{-14}$ I & $0.17 \cdot 10^{-5}$\\
 SN1006    & -41.7$^{\circ}$ & 1 & 1.21 & $0.58\cdot 10^{-14}$ 
& $0.58\cdot 10^{-14}$ & - & $0.18 \cdot 10^{-5}$\\
 Gal.Cen.& -28.9$^{\circ}$ & 0 & 0.86 & $0.48 \cdot 10^{-14}$ 
& $0.35 \cdot 10^{-14}$ & $0.95 \cdot 10^{-14}$ B& $0.15 \cdot 10^{-5}$ \\ 
 Kep1604& -21.5$^{\circ}$ & 2 & 0.82 & $1.04 \cdot 10^{-14}$ 
& $1.15 \cdot 10^{-14}$ & -  & $0.32 \cdot 10^{-5}$\\ 
 ScoXR-1   & -15.6$^{\circ}$ & 1 & 0.76 & $0.85 \cdot 10^{-14}$ 
& $0.90 \cdot 10^{-14}$ & $1.5 \cdot 10^{-14}$ B  & $0.26 \cdot 10^{-5}$\\ 
 Geminga    &  18.3$^{\circ}$ & 0 & 0.42 & $1.34 \cdot 10^{-14}$ 
& $1.17 \cdot 10^{-14}$ & $3.1 \cdot 10^{-14}$ I  & $0. 41\cdot 10^{-5}$\\
 Crab &  22.0$^{\circ}$ & 1 & 0.40 & $2.22 \cdot 10^{-14}$ 
& $2.22 \cdot 10^{-14}$ & $2.6 \cdot 10^{-14}$ B & $0.68 \cdot 10^{-5}$\\
 MRK501     &  38.8$^{\circ}$ & 0 & 0.12 & $5.40 \cdot 10^{-14}$  
& $5.44 \cdot 10^{-14}$ & - & $1.66 \cdot 10^{-5}$\\
\hline
\end{tabular}
\end{center}
%\label{tab2}
%}},%
\caption{
%\vskip 0.1 cm
%\textbf{Table 2: } 
$\mu$ flux 
limits for some sources ($90\%$ c.l.) calculated 
using the classical Poissonian method ( $\mu$ flux limit 1) and the 
prescriptions in Feldman, \& Cousins, 1998 
( $\mu$ flux limit 2).
Previous best limits
(Gaisser, 1996): B is for Baksan, I for IMB. Neutrino flux limits
are given.}
%\vskip 0.6 cm
\end{table}
\begin{tabwindow}[1,r,%
{\mbox{
\begin{tabular}{|ccc|}\hline
$E_{\nu}$ (GeV) & $P_{\nu \rightarrow \mu^{-}}$ &
$P_{\bar{\nu} \rightarrow \mu^{+}}$ \\ \hline
10      & $1.27 \times 10^{-10}$ & $9.25 \times 10^{-11}$\\
10$^{2}$& $9.73 \times 10^{-9}$ & $6.68 \times 10^{-9}$\\
10$^{3}$& $5.99 \times 10^{-7}$ & $4.12 \times 10^{-7}$ \\
10$^{4}$& $1.56 \times 10^{-5}$ & $1.14 \times 10^{-5}$\\
10$^{5}$& $1.39 \times 10^{-4}$ & $1.21 \times 10^{-4}$\\ \hline 
\end{tabular}
%\label{tab3}
}},%
{Probabilities for $\nu$s and $\bar{\nu}$s with energy $E_{\nu}$ 
to produce a $\mu$ with $E_{\mu} \ge 1$ GeV.}] 
The sample used for this analysis is larger than the one used for
the neutrino oscillation analysis (Ronga et al., 1999) because we remove the
requirement that 2 m of absorber are crossed in the lower part of the 
MACRO and we include a period in which MACRO was under construction. 
In fact, when calculating upper limits, the benefit of
increasing the exposure offsets the slight increase of the background.   
We look for statistically significant excesses of upward-going 
muons in the direction of known sources (a list we have compiled
of 40 selected sources, 
129 Egret sources (Thompson et al, 1995), 220 SNRs 
(Green, 1998), 7 sources with $\gamma$ emission above 1 TeV, 2328 GRBs in the
BATSE Catalogue (Meegan et al., 1997)) or around the direction of any of the 
detected neutrino events. 
For this directional search it is important to consider the angular 
spread between the detected $\mu$ and the parent $\nu$ due to 
the $\nu$ spectrum which determines the kinematics of the 
charged current interaction, 
the $\mu$ propagation from production to detection and the angular
resolution of the apparatus.
In Tab. 1 we show the fraction of events
accepted in a cone of $3^{\circ}$ for various differential 
$\nu$ spectral indices $\gamma$ and muon directions.
We have considered cones with half- widths of 
1.5$^{\circ}$,3$^{\circ}$,5$^{\circ}$ and 10$^{\circ}$  
around the direction of known sources or of the detected upward-going
muons. The expected atmospheric $\nu$ background is calculated by mixing
100 times the local coordinates and times of the detected upward-going 
$\mu$s in declination bands of $\pm 5^{\circ}$ around the source declination.
We find 111 clusters of $\ge 3$ events and we expect 114  
clusters from atmospheric $\nu$s in a 3$^{\circ}$ 
cone (see Fig. 1 (a)). For the 40 selected sources we find 10 
sources with $\ge 2$ events and we expect 11 in a 
3$^{\circ}$ cone.
Muon flux limits for 40 selected sources for $3^{\circ}$ cones  
are calculated using the correction factors in Tab. 1 for $\gamma = 2.1$. 
MACRO area is calculated using a GEANT-based 
%(Brun et al., 1987) 
full simulation of the detector described in 
Ambrosio et al., 1998, and it is shown as a function of declination 
in Fig. 1 (b).
We have derived flux limits (some of them are given in Tab. 2) 
calculating the upper limit from the
Poissonian probability for processes with background 
(classical method) (Barnett et al., 1996). 
\end{tabwindow}
\begin{figwindow}[1,r,%
{\mbox{
%\begin{figure}                   
\begin{tabular}{cc}
\epsfig{file=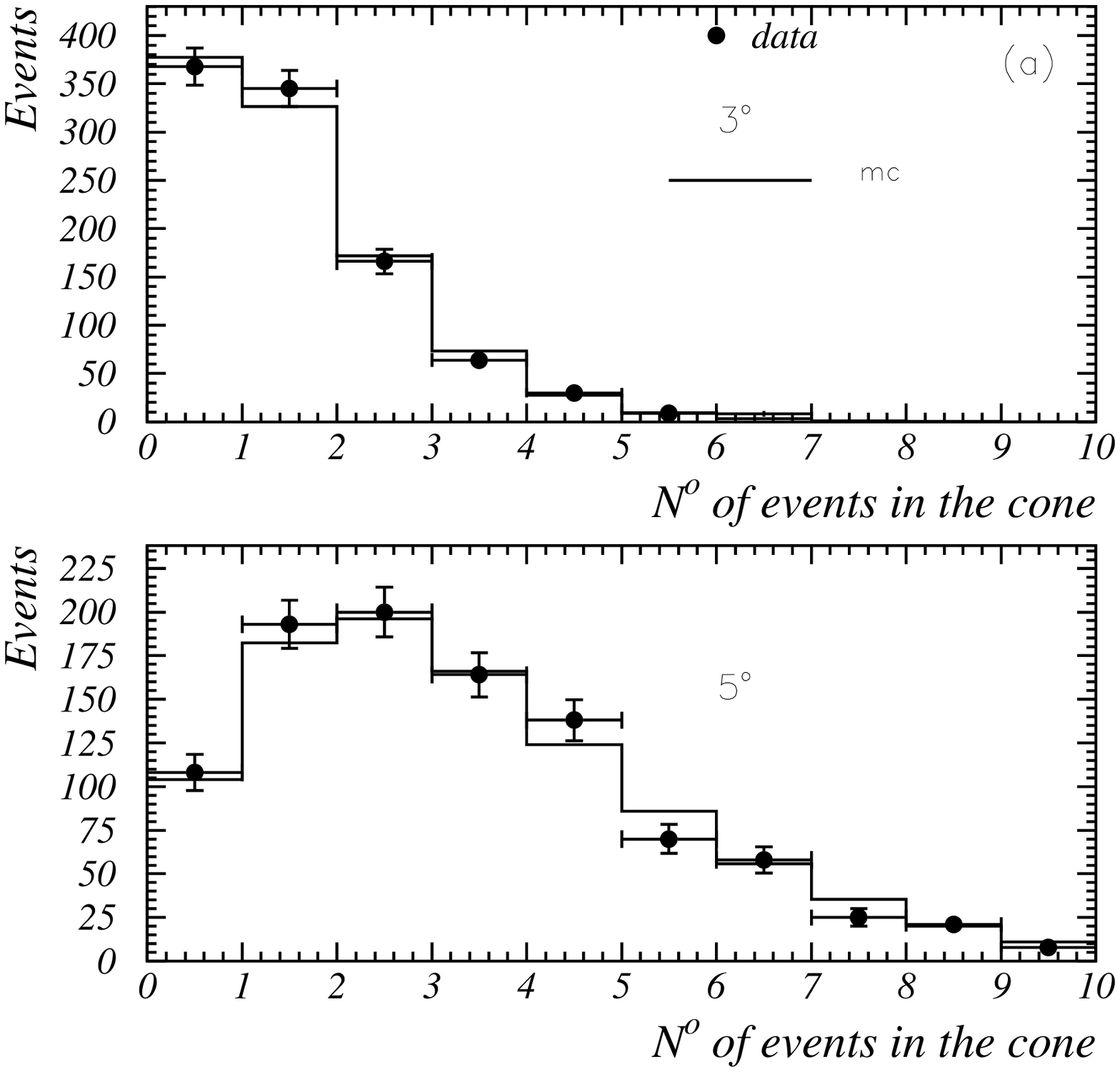,width=8.cm,height=8.5cm}&
\epsfig{file=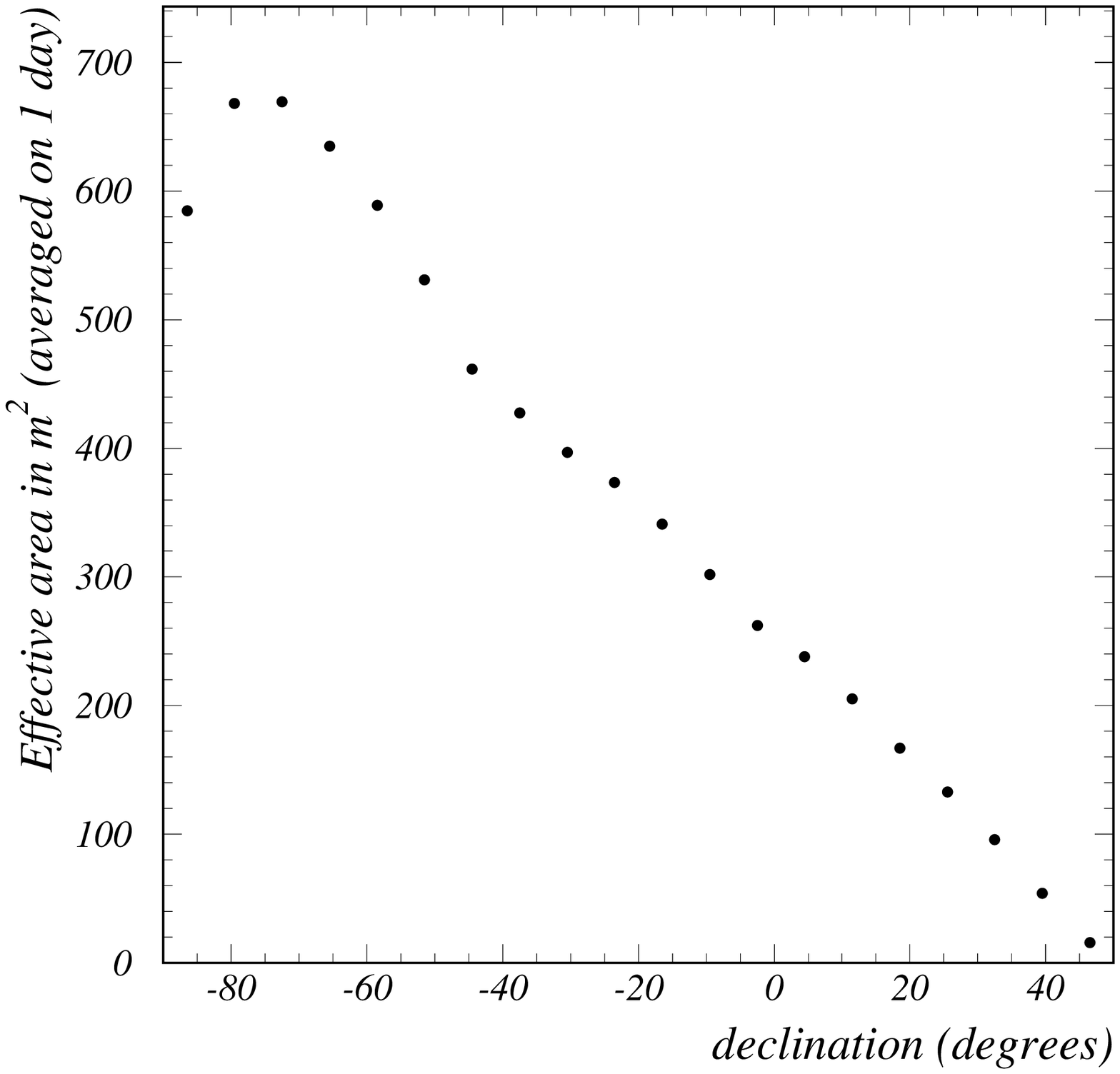,width=8.cm,height=8.5cm}
\end{tabular}
}},%
{(a) Distribution of the number of events inside cones of 
$3^{\circ}$ and $5^{\circ}$. 
(b) MACRO effective 
area as a function of declination averaged on 1 day for high energy
events ($E_{\mu} \ge 10$ GeV).}]
%\label{fig1}
%\end{figure}
The corresponding $\nu$ flux limits are obtained
considering the conversion probability of $\nu_{\mu}$ and $\bar{\nu}_{\mu}$
into muons for an energy threshold of 1 GeV (see Tab. 3).
Moreover, in Tab. 2 we compare these flux limits with that obtained 
with a recent method suggested in Feldman \& Cousins, 1998 now 
preferred by the Data Particle Book (Caso et al., 1998).
\end{figwindow}
\section{Coincidences between Gamma-ray Bursts and Upward-going Muons:}
\label{sec:gamma}
We search for correlations between 990 MACRO upward-going muons and
2328 GRBs in the BATSE 3B and 4B Catalogues (Meegan et al., 1997) 
detected from 21 Apr. 91 to 21 Feb. 99 (see Fig. 2(a)).
Considering the BATSE angular accuracy we estimate that a half-cone of 
$10^{\circ}$ includes 96.9$\%$ of the neutrinos from GRBs.
We estimate that the fraction of signal lost due to the 
angular spread between the $\mu$ and the parent $\nu$ 
in a cone $\ge 10^{\circ}$ is negligible.  
The area for upgoing muon detection in the direction of the GRBs averaged over
all of them is 118 m$^2$ (this value is small because MACRO is sensitive to
neutrinos coming from the lower hemisphere only and because during 91-94
MACRO was not complete).
We find no statistically significant correlation between MACRO and BATSE 
events considering their direction and time of detection.
As shown in Fig. 2(b), we find no $\nu$ 
events in a window of 10$^{\circ}$ from GRB directions and inside a 
time interval of $\pm 200$ s from its detection.
We have examined the event measured after 39.4 s from the BATSE event 
of 22 Sep. 95 and angular separation of 17.6$^{\circ}$. This separation is
much larger than the BATSE positional error box of 3.86$^{\circ}$ around the 
GRB.
The expected number of atmospheric $\nu$ background events 
in these windows is calculated using the delayed coincidence technique
and it is 0.04 for 10$^{\circ}$ 
%and 0.09 for 20$^{\circ}$ 
and $\Delta t = \pm 200$ s.
The corresponding upper limit (90$\%$ c.l.) on the upward-going
$\mu$ flux is 0.84 
%and 1.41 
in units of $10^{-9}$ upward-going $\mu$s cm$^{-2}$ per burst.
This limit excludes an extreme cosmic string-type scenario in 
Halzen \& Jaczko, 1996 which would produce 10$^{-1}$ $\mu$ cm$^{-2}$
for $\nu$ energies of some tens of TeV,
while it does not exclude the fireball scenario model in Bahcall \& 
Waxman, 1999.
%\begin{figure}
\begin{center}
\begin{tabular}{cc}
\epsfig{file=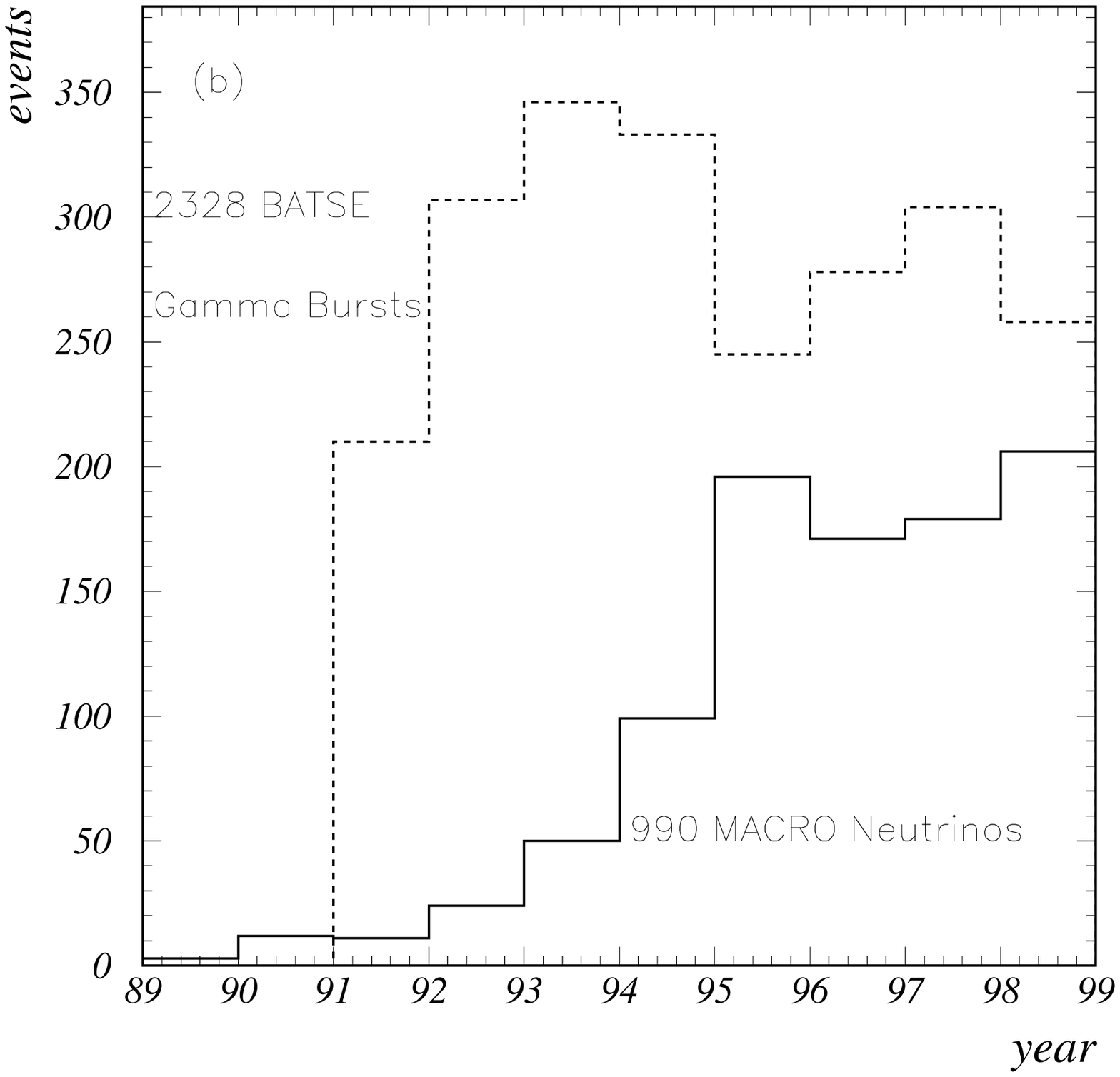,width=8.cm,height=8.5cm}&
\epsfig{file=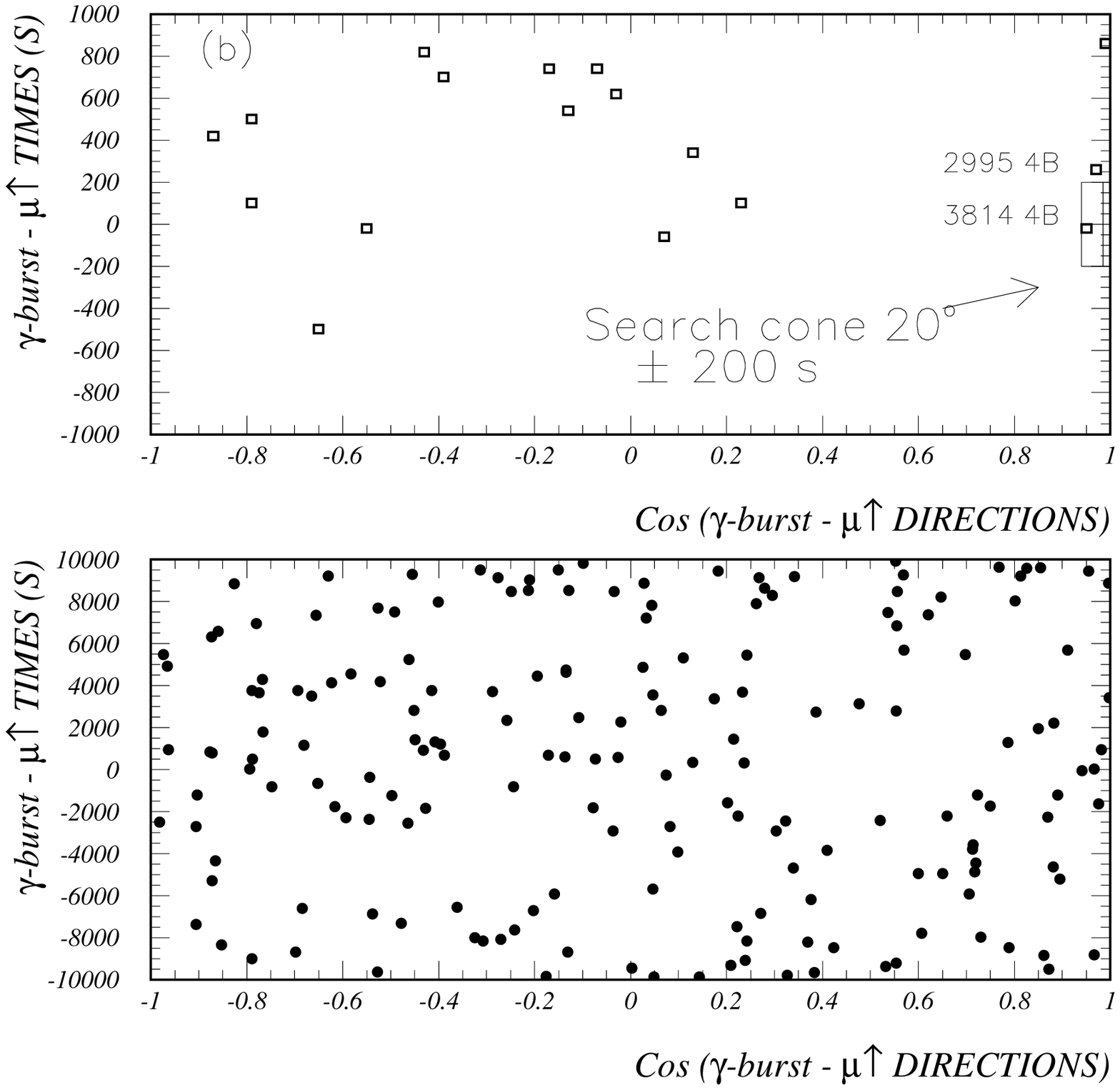,width=8.cm,height=8.5cm}
\end{tabular}
\end{center}
%\caption{
\textbf{Figure 2: } (a) Upward-going $\mu$s and BATSE GRBs vs year. (b) 
Difference in detection times vs cosine of the angular separation 
between the BATSE GRBs and upward-going $\mu$s. The 2 plots correspond to
different time scales. The $\pm 200$ s and 10$^{\circ}$-20$^{\circ}$ 
windows are indicated in the plot on the top.
%}
%\label{fig3}
%\end{figure}
\vspace{1ex}
\begin{center}
{\Large\bf References}
\end{center}
%
%  ***INSTRUCTIONS:***  Enter your references alphabetically following the 
% format
%                       of the example citations below.
Ahlen, S., et al., MACRO Collaboration, 1995, Phys. Lett. B357, 481\\
Ambrosio, M., et al., MACRO Collaboration, 1999, Phys. Rev. D59, 012003\\
Ambrosio, M., et al., MACRO Collaboration, 1998, Phys. Lett. B434, 451\\
Bahcall, J. \& Waxman, E., 1999, Phys. Rev. D59, 023002 and
1997, Phys.Rev.Lett 78, 2292\\
Barnett, R.M., et al , 1996, Review of Particle Physics,  Phys. Rev. D54, 1\\
%Brun, R., et al, 1987, CERN report DD/EE/84-1\\
Caso, C., et al, 1998, Review of Particle Physics, Z. Phys. C3, 1\\
Feldman, G.J. \& Cousins, R.D., 1998, Phys. Rev. D57, 3873\\ 
Gaisser, T.K., Halzen, F. \& Stanev, T., 1995, Phys. Rep. 258, 173\\
Gaisser, T.K., 1996, Nucl. Phys. B 48, 405\\
Galama et al., 1999, astro-ph/9903021, to appear in Nature\\
Green, D.A., 1998, http://www.mrao.cam.ac.uk/surveys/snrs/\\ 
Halzen, F. \& Jaczko, G., 1996, Phys. Rev. D54, 2779\\
Mannheim, K., Protheroe, R.J. \& Rachen, J.P., 1998, astro-ph/9812398\\
Meegan, C.A. et al., 1997, http://www.batse.msfc.nasa.gov/data/grb/catalog\\
Protheroe, R.J., Bednarek, W. \& Luo, Q., 1998, Astrop. Phys. 9, 1\\
Ronga, F. for the MACRO Collaboration, 
these Proc. 26th ICRC (Salt Lake City, 1999), HE.4.1.07\\
Thompson, D.J. et al., 1995, 
http://cossc.gsfc.nasa.gov/cossc/egret/egretcatalog\\ 
%\end{figwindow}
\end{document}